\documentclass[twocolumn,showpacs,prb]{revtex4}

\usepackage{graphicx}
\usepackage{dcolumn}
\usepackage{bm}

\begin{document}

\title{Analysis of resonant inelastic x-ray scattering in La$_2$CuO$_4$}
\author{Takuji Nomura}
\email{nomurat@spring8.or.jp}
\affiliation{
Synchrotron Radiation Research Center, 
Japan Atomic Energy Research Institute, Hyogo 679-5148, Japan}
\author{Jun-ichi Igarashi}
\affiliation{
Department of Mathematical Sciences, 
Ibaraki University, Ibaraki 310-8512, Japan}

\date{\today}

\begin{abstract}
We provide a semiquantitative explanation of a recent experiment 
on the resonant inelastic x-ray scattering (RIXS) 
in insulating cuprate La$_2$CuO$_4$
(Y.J. Kim et al., Phys. Rev. Lett. {\bf 89}, 177003 (2002).). 
We show theoretically that there are three characteristic peaks 
in RIXS spectra, two of which are attributed 
to the charge transfer excitation 
and are reasonably assigned to those observed experimentally. 
The lowest energy peak has a relatively large dispersion ($\sim 0.8$ eV)
and is suppressed near the zone corner $(\pi, \pi)$, 
in agreement with the experiment. 
We stress that electron correlation is an essential factor 
for explaining the overall energy-momentum dependence 
of the RIXS spectra consistently. 
\end{abstract}

\maketitle

\section{INTRODUCTION}

Recently it has been revealed that resonant inelastic 
x-ray scattering (RIXS) is a powerful tool 
for elucidating the electronic properties of solids~\cite{Kotani2001}. 
RIXS is a unique technique of detecting the relatively 
high-energy (of the order of 1 eV) charge excitations 
in solids. 
It is naturally expected that excitation spectra 
in such a high-energy regime reflect 
not only detailed electronic structures over a wide energy range 
but also electron correlations originating 
from strong electron-electron Coulomb repulsion. 
However, unfortunately, there are still only a few 
theoretical works on analyzing the effects 
of electron correlations on such high-energy 
excitation properties. 

\begin{figure}
\includegraphics[width=0.8\linewidth]{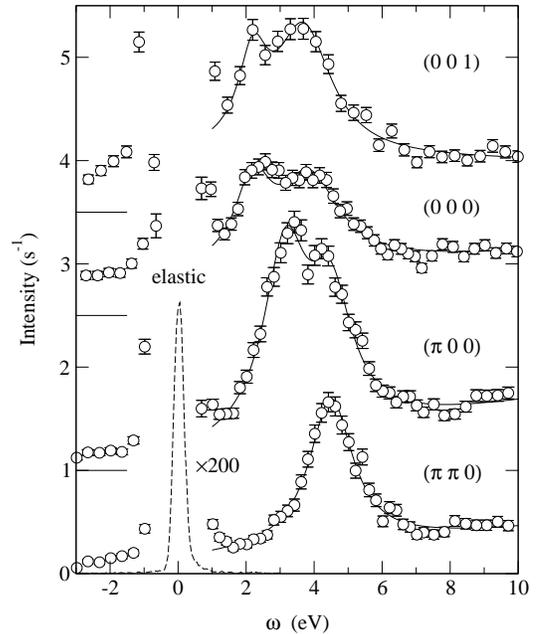}
\caption{
RIXS spectra obtained by Kim et al. 
are shown for various momentum transfers 
as a function of energy transfer. 
(Reproduced from Ref.~\onlinecite{Kim2002}.)
\label{Fig:Kim'sdata}}
\end{figure}

RIXS measurement in the hard x-ray regime has been applied 
to search for charge excitation modes 
and to determine the momentum dependence of charge excitation 
in several transition metal oxides, 
with the incident photon energy tuned to the absorption 
$K$ edge~\cite{Kao1996,Hill1998,Hasan2000,Kim2002,Inami2003}. 
A number of insulating cuprates have been investigated up to now, 
under the expectation that such a kind of excitations
might be related to the unconventional high-$T_{\rm c}$ superconductivity. 
For example, Kim et al. recently reported a detailed 
study of the momentum dependence 
of the charge transfer (CT) excitation spectra 
in La$_2$CuO$_4$ (Fig.~\ref{Fig:Kim'sdata})~\cite{Kim2002}. 
They observed two characteristic peaks in the spectra at 
around $\omega \sim 2.2$ eV and $\sim 3.9$ eV at the zone center, 
by setting the incident photon energy to the Cu absorption $K$ edge. 
The low-energy peak at 2.2 eV shifts to $\omega \sim 3$ eV 
around $(\pi,0)$, and becomes very small 
around the zone corner $(\pi, \pi)$. 
They considered that the two peaks with such a highly dispersive behavior 
are attributable to some kind of bound excitons which are formed as a result 
of Coulomb interaction, since such a bound exciton could move 
without disturbing the antiferromagnetically ordered background. 

In this article, we provide an explanation of 
the RIXS spectra for the typical antiferromagnetic
charge transfer insulator La$_2$CuO$_4$ at a microscopic scale, 
by calculating the transition probability in the RIXS process. 
In order to investigate electron correlation effects in the RIXS process, 
we use the perturbative method developed by Platzman 
and Isaacs~\cite{Platzman1998}. 
The scattering probability is calculated in the manner adopted 
by Nozi\`{e}res and Abrahams to discuss Fermi edge singularity 
in metals~\cite{Nozieres1974}. 
We use the Hartree-Fock (HF) theory 
to describe the antiferromagnetic ground state, 
and take account of the electron correlations 
in the scattering process within the random phase approximation (RPA). 
The perturbative method for band models compares 
well with other theoretical techniques based on a local atomic picture, 
such as exact diagonalization for a small cluster~\cite{Tsutsui1999}. 
We obtain a three-peak structure in the spectra at around 2, 5 and 9 eV, 
by taking account of the O2$p$ band explicitly. 
The spectrum obtained as a function of energy loss 
shows dispersive behavior depending on the momentum transfer. 
In particular, the lowest energy spectral weight 
around 2 eV shows a relatively large dispersion ($\sim$ 0.8 eV), 
and becomes small around $(\pi, \pi)$, in agreement with the experiment. 
The two peaks at around 2 and 5 eV correspond 
to charge transfer excitations and are assigned to those observed 
in the experiment by Kim et al.~\cite{Kim2002}. 

\section{MODEL and THEORY}

We introduce the model Hamiltonian for the system in the form 
$H=H_{dp}+H_{1s-3d}+H_{1s}+H_{4p}+H_x$. 
$H_{dp}$, $H_{1s}$ and $H_{4p}$ describe 
the kinetics of Cu3$d_{x^2-y^2}$(hybridized with O2$p$ orbitals), 
Cu1$s$ and Cu4$p$ electrons, respectively. 
We take a completely flat band for the Cu1$s$ electron, 
which is well localized in the solid. 
In addition, for the Cu4$p$ electrons, 
we take a simple cosine-shaped band 
with the energy minimum at the zone center. 
In any case, no detailed structure of the Cu4$p$ band 
is required to determine the shape of RIXS spectra, 
since the factor including the dispersion function 
of the Cu4$p$ band is integrated up with respect 
to the momentum and does not affect the momentum-energy 
dependence of the spectra, as we see later. 
For $H_{dp}$, we use a two-dimensional $dp$-model: 
\begin{widetext}
\begin{eqnarray}
H_{dp} = \sum_{{\bf k}\sigma} {\epsilon}_d 
d^{\dag}_{{\bf k}\sigma}d_{{\bf k}\sigma} 
+ \sum_{{\bf k}\ell\ell'\sigma} {\zeta}_{p_{\ell}p_{\ell'}}({\bf k}) 
p^{\dag}_{{\bf k}\ell\sigma}p_{{\bf k}\ell'\sigma}
+ \sum_{{\bf k}\ell\sigma} (\xi_{dp_{\ell}}({\bf k}) 
d^{\dag}_{{\bf k}\sigma}p_{{\bf k}\ell\sigma} + {\rm h.c.})\nonumber\\
+ \frac{U_{dd}}{N} \sum_{{\bf kk'q}} 
d^{\dag}_{{\bf k'+q}\uparrow} d^{\dag}_{{\bf k-q}\downarrow} 
d_{{\bf k}\downarrow} d_{{\bf k'}\uparrow}, 
\end{eqnarray}
\end{widetext}
where $d_{{\bf k}\sigma}$ and $p_{{\bf k}\ell\sigma}$ 
($d^{\dag}_{{\bf k}\sigma}$ and $p^{\dag}_{{\bf k}\ell\sigma}$) 
are annihilation (creation) operators for Cu3$d_{x^2-y^2}$ 
and O2$p_{\ell}$ electrons ($\ell=x,y$) 
with momentum ${\bf k}$ and spin $\sigma$, respectively. 
We adopt the dispersion relations 
${\xi}_{dp_{\ell}}({\bf k}) = 2i t_{dp} \sin \frac{k_{\ell}}{2}$, 
${\zeta}_{p_{\ell} p_{\ell}}({\bf k}) = \epsilon_p = 0$, 
and ${\zeta}_{p_x p_y}({\bf k}) ={\zeta}_{p_y p_x}({\bf k}) 
= 4t_{pp}\sin \frac{k_x}{2} \sin \frac{k_y}{2}$, 
where $t_{dp}=1.3$ eV and $t_{pp}=0.65$ eV, 
which were estimated from LDA calculations~\cite{Hybertsen1989}. 
The Coulomb energy and the charge transfer energy 
in HF theory are $U_{dd} = 11$ eV 
and $\epsilon_d^{HF} = -0.7$ eV, respectively. 
Within HF theory, we obtain the antiferromagnetic ground state 
with staggered spin moment $m=0.46 \mu_{\rm B}$ 
for the present parameter set. 
$H_{1s-3d}$ gives the interaction 
between Cu1$s$ holes and Cu3$d$ electrons: 
\begin{equation}
H_{1s-3d} = \frac{V}{N} \sum_{{\bf kk'q}\sigma\sigma'} 
d^{\dag}_{{\bf k'+q}\sigma} s^{\dag}_{{\bf k-q}\sigma'}
s_{{\bf k}\sigma'} d_{{\bf k'}\sigma}, 
\end{equation}
where $V$ is the core hole potential and 
is taken to be 15 eV, and $s_{{\bf k}\sigma}$ 
($s^{\dag}_{{\bf k}\sigma}$) 
is the annihilation (creation) operator for Cu1$s$ electrons  
with momentum ${\bf k}$ and spin $\sigma$. 
$H_x$ describes the transitions between the Cu1$s$ and Cu4$p$ states, 
which involve photon absorption and inverse emission processes. 
$H_x$ is of the form 
\begin{equation}
H_x = \sum_{{\bf kq}\sigma} (w({\bf q}; {\bf k}) 
p'^{\dag}_{{\bf k+q}\sigma} s_{{\bf k}\sigma} + {\rm h.c.}),
\end{equation}
where $p'^{\dag}$ is the creation operator of the Cu4$p$ electron. 
Since the Cu1$s$ wave function is highly localized 
in the coordinate space, the wave functions of the 
Cu4$p$ band states only in the region close 
to the origin contribute to the matrix elements. 
This strong local nature results in the momentum independence 
of the matrix elements. 
In the present article, we ignore the momentum dependence 
of the matrix elements  $w({\bf q}; {\bf k})$, i.e., 
$w({\bf q}; {\bf k})=w$. 

\begin{figure}
\includegraphics[width=0.7\linewidth]{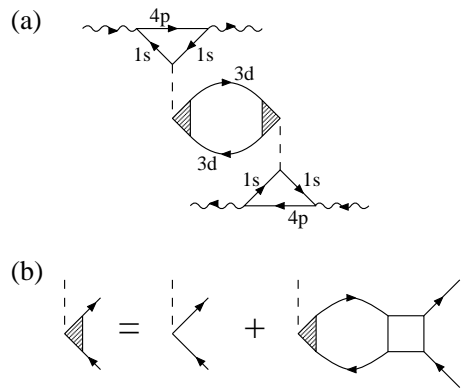}
\caption{
(a) Diagram for the transition probability in the RIXS process.
Green's functions for Cu1$s$, 4$p$ and 3$d$ 
electrons are assigned to the solid lines 
with '1s', '4p' and '3d', respectively. 
The wavy and broken lines represent the photon propagator 
and core hole potential $V$, respectively. 
The shaded triangle is the effective scattering vertex
renormalized by Coulomb interaction $U_{dd}$. 
(b) Vertex renormalization in RPA. 
The empty square represents antisymmetrized 
bare Coulomb interaction between Cu3$d$ electrons.
\label{Fig:diagram}}
\end{figure}
In the RIXS process in La$_2$CuO$_4$, 
the following scattering channel is expected to 
predominantly contribute to the main spectral weight: 
the incident x-ray with the 1$s$-4$p$ absorption energy 
excites the Cu1$s$ electron to the Cu4$p$ state. 
In the intermediate state, the created Cu1$s$ core hole 
plays the role of a localized strong scattering body 
for the Cu3$d$ electrons near the chemical potential. 
After the core hole scatters the Cu3$d$ electron system 
from the ground state to some excited states
(the most probable scattering process is the creation 
of only one electron-hole pair), 
the core hole is annihilated together with the Cu4$p$ electron
and a photon is emitted in the final state. 
Compared with the incident absorbed photon, 
the emitted photon loses momentum and energy in amounts equal 
to those transferred to the Cu3$d$-O2$p$ electrons. 
In the final state, after the Cu1$s$ core hole disappears, 
an electron-hole pair remains on the Cu3$d$-O2$p$ band. 
In the present study, we consider the lowest order 
in the core hole potential $V$, 
i.e., simple Born scattering of Cu3$d$ electrons by the core hole. 
The transition probability for the total process 
is diagrammatically expressed in Fig.~\ref{Fig:diagram}(a). 
The diagram in Fig.~\ref{Fig:diagram}(a) is converted 
straightforwardly to an analytic expression 
in the same way as described in Ref.~\onlinecite{Nozieres1974}: 
\begin{widetext}
\begin{eqnarray}
W(q_i, q_f)= (2\pi)^3 N |w|^4 \sum_{{\bf k} jj'}
\delta (E_j({\bf k})+\omega-E_{j'}({\bf k+q}))
n_j({\bf k})(1-n_{j'}({\bf k+q})) \nonumber\\
\times \biggl| \sum_{\sigma\sigma'}U_{j, d \sigma}({\bf k}) 
\Lambda_{\sigma\sigma'}(\omega,{\bf q}) 
U^{\dag}_{d \sigma', j'}({\bf k+q}) \sum_{{\bf k}_1} 
\frac{V}{(\omega_i+\epsilon_{1s}+i\Gamma_{1s}-\epsilon_{4p}({\bf k}_1))
(\omega_f+\epsilon_{1s}+i\Gamma_{1s}-\epsilon_{4p}({\bf k}_1))} 
\biggr|^2, 
\label{eq:intensity}
\end{eqnarray}
\end{widetext}
where $j^({}'{}^)$ characterizes the diagonalized band, 
$E_j({\bf k})$ is the energy dispersion of the band $j$, 
$q_{i} = (\omega_{i}, {\bf q}_{i})$ 
[$q_{f} = (\omega_{f}, {\bf q}_{f})$] 
is the energy-momentum of the initially absorbed 
[finally emitted] photon, 
and the energy and momentum transfers 
are given by $\omega=\omega_i-\omega_f$ 
and ${\bf q}={\bf q}_i-{\bf q}_f$. 
$\epsilon_{1s}$ and $\epsilon_{4p}({\bf k}_1)$ 
are the kinetic energies of the Cu1$s$ and Cu4$p$ electrons, respectively. 
$n_j({\bf k})$ is the electron occupation number 
at momentum ${\bf k}$ in energy band $j$. 
$U_{j, d \sigma}({\bf k})$ is the $(j, d \sigma)$-element 
of the diagonalization matrix of the HF Hamiltonian. 
We set the incident photon energy near the absorption edge, 
i.e., $\omega_i \approx \epsilon_{4p}(0) - \epsilon_{1s}$, 
and assume that the decay rate of the Cu1$s$ hole is $\Gamma_{1s}=0.8$ eV. 
In the present study, we simply use the vertex function 
$\Lambda_{\sigma\sigma'}(q)$ within RPA. 
The vertex renormalization within RPA 
is diagrammatically represented in Fig.~\ref{Fig:diagram}(b). 
It is interesting to note that the factor 
$n_j({\bf k}) |U_{j, d \sigma}({\bf k})|^2$ 
[$(1-n_{j'}({\bf k+q})) |U^{\dag}_{d \sigma', j'}({\bf k+q})|^2$] 
in Eq.~(\ref{eq:intensity}) represents the partial 
occupation number of the Cu3$d$ electron [hole] 
with momentum ${\bf k}$ [${\bf k+q}$] 
and spin $\sigma$ [$\sigma'$] in the band $j$ [$j'$]. 
According to our investigations, the product 
$n_j({\bf k})(1-n_{j'}({\bf k+q})) 
|U_{j, d \sigma}({\bf k}) U^{\dag}_{d \sigma', j'}({\bf k+q})|^2$ 
is the dominant factor in the scattering amplitude. 
Therefore, the RIXS spectrum in La$_2$CuO$_4$ reflects 
the Cu3$d$ partial electron and hole occupation numbers, 
that is, the Cu3$d$ partial density of states. 

\section{RESULTS}

\begin{figure}
\includegraphics[width=0.9\linewidth]{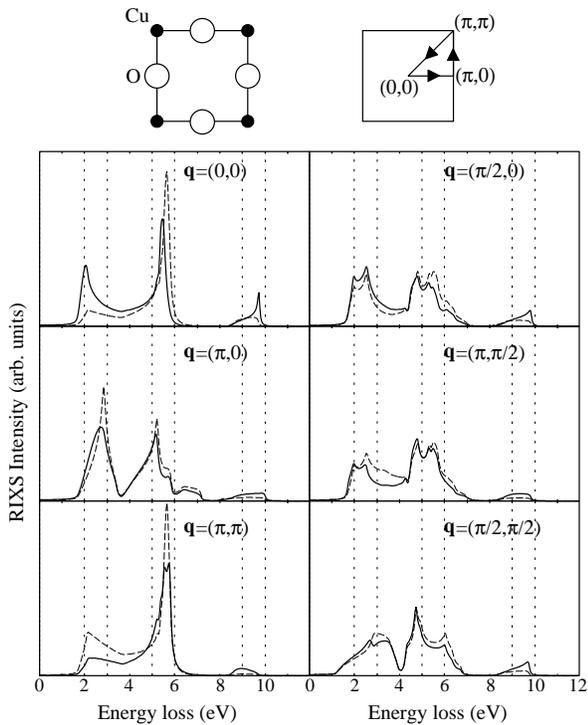}
\caption{
The upper figures show the Cu-O plaquette and the contour 
in the first Brillouin zone (BZ). 
The lower six panels show the RIXS spectra 
as a function of the energy loss 
$\omega=\omega_i-\omega_f$ on the above contour in BZ.
The thick solid (thin broken) curves are results 
obtained with (without) RPA vertex renormalization.
\label{Fig:intensity}}
\end{figure}

The numerical results of the RIXS intensity calculated 
using Eq.~(\ref{eq:intensity}) are displayed 
in Fig.~\ref{Fig:intensity}. 
We see the three characteristic peaks 
at around $\omega \sim 2, 5$ and $9$ eV. 
The lowest energy peak at 2 eV shows relatively large 
dispersive behavior. The peak shifts by approximately 0.8 eV 
at ${\bf q} \approx (\pi,0)$ and becomes small 
at ${\bf q} \approx (\pi,\pi)$. 
This characteristic momentum dependence 
is consistent with the results in Fig.~\ref{Fig:Kim'sdata}. 
The peak at 5 eV is assigned 
to the peak observed experimentally at 3.9 eV. 
The crude tight-binding fitting would be responsible 
for the 1 eV deviation of the peak position, and would still be
insufficient to reproduce the electronic structure 
in such a high-energy region. 
The two peaks at 2 and 5 eV originate from CT excitation: 
Cu3$d$ electrons hybridized with the O2$p$ band 
are excited to the upper Cu3$d$ band. 
This two-peak structure is a result of the  structure 
of the Cu3$d$ partial density of states mixed in the broad O2$p$ band. 
On the other hand, the small peak at around 9 eV 
corresponds to the transition from the Cu3$d$ lower Hubbard band 
to the Cu3$d$ upper Hubbard band. 
The transition probability for this 9 eV process is relatively low. 
This is because the majority spin state 
at a local Cu site is almost fully occupied by one electron, 
and the hole occupation number in that state is almost zero. 
The actual localized moment will be almost $1\mu_{\rm B}$, 
since the electron occupation number per one Cu site is almost one. 
As we have mentioned above, the intensity is closely related 
to the product of the Cu3$d$ electron occupation number 
and the Cu3$d$ hole occupation number. 
In the local picture, this product 
is expressed as $n_{d,i\sigma}(1-n_{d,i\sigma})$ 
($n_{d,i\sigma}$ is Cu3$d$ electron number at Cu site $i$ 
with spin $\sigma$), and becomes almost zero 
in the limit of $m=1\mu_{\rm B}$. 
In the present HF calculation, we may have underestimated 
the electron occupation number in the majority spin state at a Cu site. 
In other words the magnetic moment $m=0.46\mu_{\rm B}$ obtained 
in the HF theory might be smaller than the actual localized moment. 
Therefore, we should consider that the present method still 
overestimates the transition probability for this 9 eV process. 
Actually, the intensity for this process would 
be too small to be observed. 
Therefore this 9 eV weight should not be identified 
with the 7.2 eV peak observed 
in the experiment (peak c in Ref.~\onlinecite{Kim2002}), 
whose origin should be discussed on the basis 
of other scenarios. 

\begin{figure}
\includegraphics[width=0.95\linewidth]{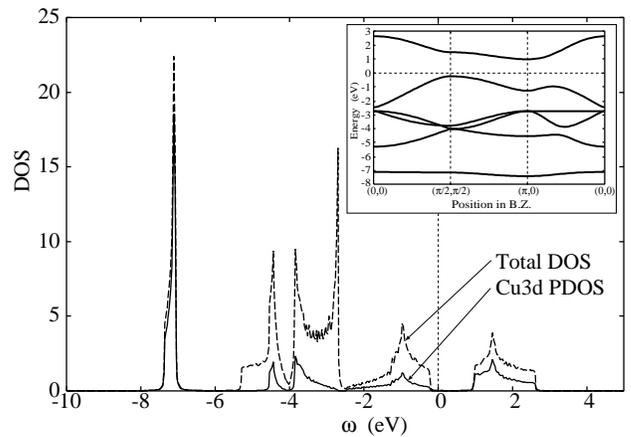}
\caption{
The total density of states and the Cu3$d_{x^2-y^2}$ partial 
density of states are shown by the solid and broken lines, respectively. 
The inset shows the band dispersion of the model in HF theory.}
\label{Fig:dpdos}
\end{figure}

Here we inspect the electronic structure 
and the density of states (DOS) of the model 
in the antiferromagnetic ground state. 
The Cu3$d$ partial density of states (PDOS) 
and the total density of sates 
are displayed in Fig.~\ref{Fig:dpdos}. 
The 2 and 5 eV RIXS peaks are assigned 
to the electronic excitation processes 
from the Cu3$d$ PDOS weights 
around $\omega \approx -1$ and $-4$ (eV) 
to the upper Cu3$d$ band 
around $\omega \approx 2$ (eV), respectively. 
These Cu3$d$ PDOS weights 
around $\omega \approx -1$ and $-4$ (eV) 
originate from the strong hybridization 
of the Cu3$d$ and O2$p$ orbitals. 
Therefore the two peak structures are attributed 
to Cu3$d$-O2$p$ CT excitations. 

It is of interest to discuss the effect of the correlations 
on the shape of the spectra. In Fig.~\ref{Fig:intensity}, 
we compare the RPA-corrected case with the uncorrected 
case (i.e., zero order in $U_{dd}$: $\Lambda_{\sigma\sigma'}(q) 
\rightarrow \delta_{\sigma\sigma'}$ in Eq.~(\ref{eq:intensity})). 
We find that the lowest order uncorrected calculation 
reproduces roughly the overall structure of the RIXS spectra. 
This is because the spectral weight is determined mainly 
by the partial occupation number of Cu3$d$ electrons in each band, 
and the detailed scattering process is not necessary 
in the first step of analysis. 
However, the low energy peak structure at around 2 eV 
is modified somewhat by the RPA corrections. 
The correlation effects between Cu3$d$ electrons 
cause an anisotropy and modify 
the simple $s$-wave scattering amplitude, 
and result in the enhancement and suppression 
of the spectral weight at ${\bf q}=(0,0)$ 
and ${\bf q}=(\pi,\pi)$, respectively. 

\begin{figure}
\includegraphics[width=0.9\linewidth]{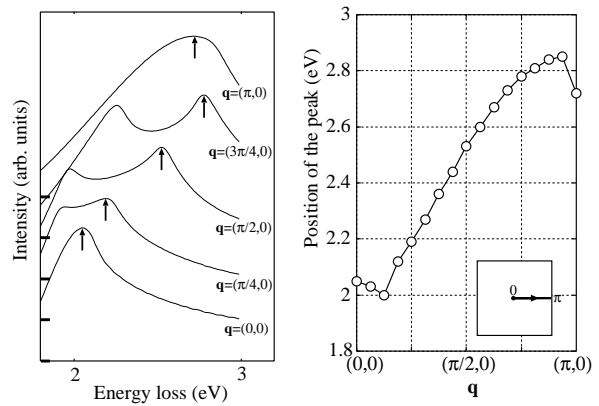}
\caption{
Left panel shows details of the spectra along the line $(0,0)$-$(\pi,0)$ 
between $\omega=2$-3 eV. 
The data are offset vertically, and the thick horizontal bars drawn 
on the left vertical axis denote the baseline for each spectrum. 
The arrows denote the positions of the main peak. 
Right panel shows the position of the main peak
as a function of ${\bf q}$.
\label{Fig:dispersion}}
\end{figure}

We focus our attention on the dispersive behavior 
of the low-energy peak at 2 eV. 
We display the detailed behaviors of the 2 eV peak 
in Fig.~\ref{Fig:dispersion}, along the line $(0,0)$-$(\pi,0)$. 
We find that there is a fine structure in the 2 eV peak: 
the peak splits into two peaks. 
Such a fine structure is not observed in the experiment
~\cite{Kim2002}, maybe due to the low resolution. 
The shift of the peak between the points $(0,0)$ and $(\pi,0)$
is about 0.8 eV. 
This is slightly smaller than the experimental result. 
We consider that this deviation originates 
from our underestimation of the O2$p$ bandwidth. 

\section{DISCUSSION}

We have used the LDA parameters for the model. 
Although there are still much room for discussing 
the validity of the application 
of the LDA parameters to excitation processes, 
the parameters would provide a good starting point.  
To be more quantitative, the renormalization effects 
due to correlations should be taken into account 
in the future. 

We briefly note some effects of other factors which 
are not included in the above calculations: effects of 
on-site Coulomb interaction $U_{pp}$ between O2$p$ electrons, 
inter-site Coulomb interaction $U_{dp}$ 
between O2$p$ and Cu3$d$ electrons, 
higher orders in the core hole potential $V$, 
self-energy corrections and 
Coulomb interaction $U_{3d-4p}$ 
between Cu3$d$ and 4$p$ electrons, 
multi-pair creation, phonons. 
We expect that these effects are not crucial 
in explaining the observed spectral properties. 

We employed multiband RPA including $U_{pp}$, 
but found that $U_{pp}$ only reinforces the effect 
of the Coulomb interaction $U_{dd}$. 
We cannot exclude the possibility that $U_{dp}$ 
modifies the analytic properties of the spectral functions, 
i.e., novel poles of the spectral function could appear 
on the complex $\omega$ plane near the real axis. 
In such a case, some peak structures 
related to a bound exciton mode would be added to the spectra. 
However, note that there is no necessity for the additional peak 
to be located around the observed positions at 2 and 4 eV. 
Such a novel structure would not be consistent 
with the experiment. 

We could investigate the effects of higher orders of $V$ 
by summing up the perturbation terms to the infinite order, 
similarly to $T$-matrix approximation for single impurity problems. 
We speculate from our preliminary work 
that the higher orders in $V$ only negligibly
modify the spectral shape, although they modify 
the absolute intensity of the spectra. 

Concerning the self-energy corrections, 
detailed investigations are a future work. 
In x-ray absorption process, 
$U_{3d-4p}$ could play an important role 
for ``Extrinsic losses'' (e.g., Ref.~\onlinecite{Rehr2003})
through self-energy corrections, 
leaving electron-hole pairs on the $3d$ band 
in the final state. 
But, $U_{3d-4p}$ would not be so essential 
in the present RIXS process, 
since the $4p$ electron is annihilated rapidly 
with Cu$1s$ hole in the final state of the RIXS. 

Multi-pair creation processes may become 
more important for metals, leading 
to the Fermi edge singularity in the spectra. 
However multi-pair creation processes
are not so essential in insulators, 
since they require much excitation energy 
(the energy per one pair is given 
by the insulating gap, about 2eV in the present case). 

Regarding phonon effect, phonons can be excited 
in order to screen the core hole potential 
in the intermediate state. 
However, in the present RIXS, the phonon process cannot be 
the dominant process to screen the core hole potential, 
since the annihilation of the core hole 
is much faster than the screening process by phonons.
\\
\section{CONCLUSION}

In conclusion, we note the following points. 
(i) We have not employed the physical picture 
that a kind of bound exciton formed by Coulomb interaction 
appears and moves in the CuO$_2$ plane. 
The two-peak structure of the observed spectra 
is attributable to the structure 
of the Cu3$d$ partial density of states mixed 
with the O2$p$ band, rather than some kind of bound exciton mode. 
(ii) The present formulation by the perturbative method 
is applicable to more complex and realistic electronic structures, 
in contrast to small cluster calculations 
and exact diagonalization techniques, 
whose applicability is limited to simple solids consisting 
of a few kinds of elements. 

\begin{acknowledgments}
The authors are grateful to Dr. Y.J. Kim for the discussions. 
The numerical computations were partly performed 
at the Yukawa Institute Computer Facility in Kyoto University. 
\end{acknowledgments}

\bibliography{rixs}

\begin{thebibliography}{11}
\expandafter\ifx\csname natexlab\endcsname\relax\def\natexlab#1{#1}\fi
\expandafter\ifx\csname bibnamefont\endcsname\relax
  \def\bibnamefont#1{#1}\fi
\expandafter\ifx\csname bibfnamefont\endcsname\relax
  \def\bibfnamefont#1{#1}\fi
\expandafter\ifx\csname citenamefont\endcsname\relax
  \def\citenamefont#1{#1}\fi
\expandafter\ifx\csname url\endcsname\relax
  \def\url#1{\texttt{#1}}\fi
\expandafter\ifx\csname urlprefix\endcsname\relax\def\urlprefix{URL }\fi
\providecommand{\bibinfo}[2]{#2}
\providecommand{\eprint}[2][]{\url{#2}}

\bibitem[{\citenamefont{Kotani and Shin}(2001)}]{Kotani2001}
\bibinfo{author}{\bibfnamefont{A.}~\bibnamefont{Kotani}} \bibnamefont{and}
  \bibinfo{author}{\bibfnamefont{S.}~\bibnamefont{Shin}},
  \bibinfo{journal}{Rev.\ Mod.\ Phys.} \textbf{\bibinfo{volume}{73}},
  \bibinfo{pages}{203} (\bibinfo{year}{2001}).

\bibitem[{\citenamefont{Kao et~al.}(1996)\citenamefont{Kao, Caliebe, Hastings,
  and Gillet}}]{Kao1996}
\bibinfo{author}{\bibfnamefont{C.-C.} \bibnamefont{Kao}},
  \bibinfo{author}{\bibfnamefont{W.}~\bibnamefont{Caliebe}},
  \bibinfo{author}{\bibfnamefont{J.}~\bibnamefont{Hastings}}, \bibnamefont{and}
  \bibinfo{author}{\bibfnamefont{J.-M.} \bibnamefont{Gillet}},
  \bibinfo{journal}{Phys.\ Rev.\ B} \textbf{\bibinfo{volume}{54}},
  \bibinfo{pages}{16361} (\bibinfo{year}{1996}).

\bibitem[{\citenamefont{Hill et~al.}(1998)\citenamefont{Hill, Kao, Caliebe,
  Matsubara, Kotani, Peng, and Greene}}]{Hill1998}
\bibinfo{author}{\bibfnamefont{J.}~\bibnamefont{Hill}},
  \bibinfo{author}{\bibfnamefont{C.-C.} \bibnamefont{Kao}},
  \bibinfo{author}{\bibfnamefont{W.}~\bibnamefont{Caliebe}},
  \bibinfo{author}{\bibfnamefont{M.}~\bibnamefont{Matsubara}},
  \bibinfo{author}{\bibfnamefont{A.}~\bibnamefont{Kotani}},
  \bibinfo{author}{\bibfnamefont{J.}~\bibnamefont{Peng}}, \bibnamefont{and}
  \bibinfo{author}{\bibfnamefont{R.}~\bibnamefont{Greene}},
  \bibinfo{journal}{Phys.\ Rev.\ Lett.} \textbf{\bibinfo{volume}{80}},
  \bibinfo{pages}{4967} (\bibinfo{year}{1998}).

\bibitem[{\citenamefont{Hasan et~al.}(2000)\citenamefont{Hasan, Isaacs, Shen,
  L.L.Miller, Tsutsui, Tohyama, and Maekawa}}]{Hasan2000}
\bibinfo{author}{\bibfnamefont{M.}~\bibnamefont{Hasan}},
  \bibinfo{author}{\bibfnamefont{E.}~\bibnamefont{Isaacs}},
  \bibinfo{author}{\bibfnamefont{Z.-X.} \bibnamefont{Shen}},
  \bibinfo{author}{\bibnamefont{L.L.Miller}},
  \bibinfo{author}{\bibfnamefont{K.}~\bibnamefont{Tsutsui}},
  \bibinfo{author}{\bibfnamefont{T.}~\bibnamefont{Tohyama}}, \bibnamefont{and}
  \bibinfo{author}{\bibfnamefont{S.}~\bibnamefont{Maekawa}},
  \bibinfo{journal}{Science} \textbf{\bibinfo{volume}{288}},
  \bibinfo{pages}{1811} (\bibinfo{year}{2000}).

\bibitem[{\citenamefont{Kim et~al.}(2002)\citenamefont{Kim, Hill, Burns,
  Wakimoto, Birgeneau, Casa, Cog, and Venkataraman}}]{Kim2002}
\bibinfo{author}{\bibfnamefont{Y.}~\bibnamefont{Kim}},
  \bibinfo{author}{\bibfnamefont{J.}~\bibnamefont{Hill}},
  \bibinfo{author}{\bibfnamefont{C.}~\bibnamefont{Burns}},
  \bibinfo{author}{\bibfnamefont{S.}~\bibnamefont{Wakimoto}},
  \bibinfo{author}{\bibfnamefont{R.}~\bibnamefont{Birgeneau}},
  \bibinfo{author}{\bibfnamefont{D.}~\bibnamefont{Casa}},
  \bibinfo{author}{\bibfnamefont{T.}~\bibnamefont{Cog}}, \bibnamefont{and}
  \bibinfo{author}{\bibfnamefont{C.}~\bibnamefont{Venkataraman}},
  \bibinfo{journal}{Phys.\ Rev.\ Lett.} \textbf{\bibinfo{volume}{89}},
  \bibinfo{pages}{177003} (\bibinfo{year}{2002}).

\bibitem[{\citenamefont{Inami et~al.}(2003)\citenamefont{Inami, Fukuda, Mizuki,
  Ishihara, Kondo, Nakao, Matsumura, Hirota, Murakami, Maekawa
  et~al.}}]{Inami2003}
\bibinfo{author}{\bibfnamefont{T.}~\bibnamefont{Inami}},
  \bibinfo{author}{\bibfnamefont{T.}~\bibnamefont{Fukuda}},
  \bibinfo{author}{\bibfnamefont{J.}~\bibnamefont{Mizuki}},
  \bibinfo{author}{\bibfnamefont{S.}~\bibnamefont{Ishihara}},
  \bibinfo{author}{\bibfnamefont{H.}~\bibnamefont{Kondo}},
  \bibinfo{author}{\bibfnamefont{H.}~\bibnamefont{Nakao}},
  \bibinfo{author}{\bibfnamefont{T.}~\bibnamefont{Matsumura}},
  \bibinfo{author}{\bibfnamefont{K.}~\bibnamefont{Hirota}},
  \bibinfo{author}{\bibfnamefont{Y.}~\bibnamefont{Murakami}},
  \bibinfo{author}{\bibfnamefont{S.}~\bibnamefont{Maekawa}},
  \bibnamefont{et~al.}, \bibinfo{journal}{Phys.\ Rev.\ B}
  \textbf{\bibinfo{volume}{67}}, \bibinfo{pages}{045108}
  (\bibinfo{year}{2003}).

\bibitem[{\citenamefont{Platzman and Isaacs}(1998)}]{Platzman1998}
\bibinfo{author}{\bibfnamefont{P.}~\bibnamefont{Platzman}} \bibnamefont{and}
  \bibinfo{author}{\bibfnamefont{E.}~\bibnamefont{Isaacs}},
  \bibinfo{journal}{Phys.\ Rev.\ B} \textbf{\bibinfo{volume}{57}},
  \bibinfo{pages}{11107} (\bibinfo{year}{1998}).

\bibitem[{\citenamefont{Nozi\`{e}res and Abrahams}(1974)}]{Nozieres1974}
\bibinfo{author}{\bibfnamefont{P.}~\bibnamefont{Nozi\`{e}res}}
  \bibnamefont{and} \bibinfo{author}{\bibfnamefont{E.}~\bibnamefont{Abrahams}},
  \bibinfo{journal}{Phys.\ Rev.\ B} \textbf{\bibinfo{volume}{10}},
  \bibinfo{pages}{3099} (\bibinfo{year}{1974}).

\bibitem[{\citenamefont{Tsutsui et~al.}(1999)\citenamefont{Tsutsui, Tohyama,
  and Maekawa}}]{Tsutsui1999}
\bibinfo{author}{\bibfnamefont{K.}~\bibnamefont{Tsutsui}},
  \bibinfo{author}{\bibfnamefont{T.}~\bibnamefont{Tohyama}}, \bibnamefont{and}
  \bibinfo{author}{\bibfnamefont{S.}~\bibnamefont{Maekawa}},
  \bibinfo{journal}{Phys.\ Rev.\ Lett.} \textbf{\bibinfo{volume}{83}},
  \bibinfo{pages}{3705} (\bibinfo{year}{1999}).

\bibitem[{\citenamefont{Hybertsen and Schluter}(1989)}]{Hybertsen1989}
\bibinfo{author}{\bibfnamefont{M.~S.} \bibnamefont{Hybertsen}}
  \bibnamefont{and} \bibinfo{author}{\bibfnamefont{M.}~\bibnamefont{Schluter}},
  \bibinfo{journal}{Phys.\ Rev.\ B} \textbf{\bibinfo{volume}{39}},
  \bibinfo{pages}{9028} (\bibinfo{year}{1989}).

\bibitem[{\citenamefont{Rehr}(2003)}]{Rehr2003}
\bibinfo{author}{\bibfnamefont{J.}~\bibnamefont{Rehr}},
  \bibinfo{journal}{Found.\ Phys.} \textbf{\bibinfo{volume}{33}},
  \bibinfo{pages}{1735} (\bibinfo{year}{2003}).

\end{thebibliography}

\end{document}